\documentclass[aps,prd,a4paper,twocolumn,amsmath,showpacs,superscriptaddress,nofootinbib,preprintnumbers]{revtex4-1}  
\pdfoutput=1
\usepackage{amssymb,amsmath,latexsym,mathrsfs}
\usepackage{graphicx,subfigure}
\usepackage{epsfig}
\usepackage{varioref,xr-hyper}
\usepackage{color}
\usepackage{multirow}
\usepackage{array}
\usepackage{hyperref}
\usepackage{wasysym}
\usepackage{color}
\usepackage{float}
\usepackage[utf8]{inputenc}
\usepackage[T1]{fontenc}
\usepackage{footmisc}

\newcommand{\mnu}{M_{\nu}}
\newcommand{\lcdm}{\Lambda\mathrm{CDM}}
\newcommand{\eV}{\mathrm{eV}}
\newcommand{\Lim}[1]{\raisebox{0.5ex}{\scalebox{0.8}{$\displaystyle \lim_{#1}\;$}}}

\begin{document}
\preprint{LCTP-18-03,NORDITA-2018-004,IFIC/18-02}

\title{Constraints on the sum of the neutrino masses in dynamical dark energy models with $w(z) \geq -1$ are tighter than those obtained in $\lcdm$}

\author{Sunny Vagnozzi}
\email{sunny.vagnozzi@fysik.su.se}
\affiliation{The Oskar Klein Centre for Cosmoparticle Physics, Department of Physics, Stockholm University, SE-106 91 Stockholm, Sweden}
\affiliation{The Nordic Institute for Theoretical Physics (NORDITA), Roslagstullsbacken 23, SE-106 91 Stockholm, Sweden}

\author{Suhail Dhawan}
\affiliation{The Oskar Klein Centre for Cosmoparticle Physics, Department of Physics, Stockholm University, SE-106 91 Stockholm, Sweden}

\author{Martina Gerbino}
\affiliation{The Oskar Klein Centre for Cosmoparticle Physics, Department of Physics, Stockholm University, SE-106 91 Stockholm, Sweden}

\author{Katherine Freese}
\affiliation{The Oskar Klein Centre for Cosmoparticle Physics, Department of Physics, Stockholm University, SE-106 91 Stockholm, Sweden}
\affiliation{The Nordic Institute for Theoretical Physics (NORDITA), Roslagstullsbacken 23, SE-106 91 Stockholm, Sweden}
\affiliation{Leinweber Center for Theoretical Physics, Department of Physics, University of Michigan, Ann Arbor, MI 48109, USA}

\author{Ariel Goobar}
\affiliation{The Oskar Klein Centre for Cosmoparticle Physics, Department of Physics, Stockholm University, SE-106 91 Stockholm, Sweden}

\author{Olga Mena}
\affiliation{Instituto de F\'{i}sica Corpuscular (IFIC), Universidad de Valencia-CSIC, E-46980, Valencia, Spain}

\date{\today}

\begin{abstract}
We explore cosmological constraints on the sum of the three active neutrino masses $\mnu$ in the context of dynamical dark energy (DDE) models with equation of state (EoS) parametrized as a function of redshift $z$ by  $w(z)=w_0+w_a\,z/(1+z)$, and satisfying $w(z)\geq-1$ for all $z$. We make use of Cosmic Microwave Background data from the Planck satellite, Baryon Acoustic Oscillations measurements, and Supernovae Ia luminosity distance measurements, and perform a Bayesian analysis. We show that, within these models, the bounds on $\mnu$ \textit{do not degrade} with respect to those obtained in the $\lcdm$ case; in fact the bounds are slightly tighter, despite the enlarged parameter space. We explain our results based on the observation that, for fixed choices of $w_0\,,w_a$ such that $w(z)\geq-1$ (but not $w=-1$ for all $z$), the upper limit on $\mnu$ is tighter than the $\lcdm$ limit because of the well-known degeneracy between $w$ and $M_{\nu}$. The Bayesian analysis we have carried out then integrates over the possible values of $w_0$-$w_a$ such that $w(z)\geq-1$, all of which correspond to tighter limits on $M_\nu$ than the $\lcdm$ limit. We find a 95\% credible interval (C.I.) upper bound of $\mnu<0.13\,\eV$. This bound can be compared with the 95\%~C.I. upper bounds of $\mnu<0.16\,\eV$, obtained within the $\lcdm$ model, and $\mnu<0.41\,\eV$, obtained in a DDE model with arbitrary EoS (which allows values of $w < -1$). Contrary to the results derived for DDE models with arbitrary EoS, we find that a dark energy component with $w(z)\geq-1$ is unable to alleviate the tension between high-redshift observables and direct measurements of the Hubble constant $H_0$. Finally, in light of the results of this analysis, we also discuss the implications for DDE models of a possible determination of the neutrino mass ordering by laboratory searches.
\end{abstract}

\maketitle

\section{Introduction}
\label{sec:introduction}

The nature of the dark energy (DE) driving the accelerated expansion of the Universe remains one of the greatest open problems in cosmology~\cite{Riess:1998cb,Perlmutter:1998np,Astier:2012ba,Ade:2015xua,Alam:2016hwk,Abbott:2017wau,
Sahni:2004ai,Nojiri:2006ri,Frieman:2008sn,Bamba:2012cp,Mortonson:2013zfa,Huterer:2017buf,
Nielsen:2015pga,Haridasu:2017lma,Dhawan:2017leu,Tutusaus:2017ibk,Dam:2017xqs,Andersen:2017uew,Tutusaus:2018ulu}. The most economical DE candidate is a cosmological constant (CC) related to vacuum energy density. The equation of state (EoS) of such a component is $w_{\text{DE}}=P_{\text{DE}}/\rho_{\text{DE}}=-1$, where $P_{\text{DE}}$ and $\rho_{\text{DE}}$ are the pressure and energy density of the DE respectively. The CC model is however at odds with theoretical expectations of the magnitude of the CC, an issue dubbed the \textit{cosmological constant problem}~\cite{Weinberg:1988cp,Carroll:2000fy,Peebles:2002gy,Martin:2012bt}. An alternative solution to this issue posits the existence of a dynamical dark energy (DDE) component~\cite{Peebles:1987ek,Sahni:1999gb,Copeland:2006wr}, which implies a redshift-dependent equation of state $w(z)$.
 
The value $w=-1$ plays an important role from the theoretical point of view, as it demarcates two very different physical regimes~\cite{Caldwell:1999ew,Elizalde:2004mq,Vikman:2004dc,Nojiri:2005sx,Briscese:2006xu,
Jhingan:2008ym,Nojiri:2013ru,Shafer:2013pxa,Ludwick:2015dba,Ludwick:2017tox,Barenboim:2017sjk}. The energy density of a component with $w<-1$ increases with time as the Universe expands. More importantly, such a component would violate the dominant energy condition, which imposes the inequality $\vert P \vert \leq \rho$~\cite{Hawking:1973uf}~\footnote{We note that this issue can be avoided in models in which the effective dark energy component appears to violate the dominant energy condition while the full theory does not (e.g.~\cite{Barenboim:2017sjk}).}. A component with EoS $w<-1$ is usually referred to as "phantom energy". It has been shown that a Universe dominated by a phantom DE component would end in a Big Rip: the dissociation of any bound system due to the DE energy density becoming infinite in a finite amount of time~\cite{Caldwell:2003vq}
~\footnote{A notable exception is found in the case when $w(z) \to -1$ asymptotically in the future (that is, where the future geometry is asymptotically de Sitter)~\cite{Frampton:2011sp}. 
This occurs for instance in bimetric gravity~\cite{VonStrauss:2011mq,Akrami:2015qga,Schmidt-May:2015vnx,Mortsell:2017fog} as well as in other modified gravity theories (see e.g.~\cite{Astashenok:2012tv,Myrzakulov:2013mja,Odintsov:2015zza,Oikonomou:2015qha,Odintsov:2015ynk}) or more complex dark energy scenarios~\cite{BouhmadiLopez:2004me,Cattoen:2005dx,Wei:2005fq,BouhmadiLopez:2006fu,Zhang:2009xj}.}. Examples of non-phantom dark energy models include many quintessence models~\cite{Ratra:1987rm,Caldwell:1997ii,Caldwell:2005tm,Linder:2007wa} and Cardassian cosmology~\cite{Freese:2002sq,Gondolo:2002fh,Freese:2005ff}.

Recent works have studied and forecasted cosmological constraints on the sum of the three active neutrino masses, $M_{\nu} = \sum_i m_i$ (where $m_i$ are the masses of the individual mass eigenstates), within the context of the standard cosmological model, the $\lcdm$ model, which fixes $w=-1$~\cite{Palanque-Delabrouille:2015pga,Cuesta:2015iho,Huang:2015wrx,Giusarma:2016phn,Vagnozzi:2017ovm,Ade:2015xua,Zhen:2015yba,
Gerbino:2015ixa,DiValentino:2015wba,Allison:2015qca,Moresco:2016nqq,Alam:2016hwk,Oh:2016wls,Archidiacono:2016lnv,Capozzi:2017ipn,
Couchot:2017pvz,Caldwell:2017mqu,Doux:2017tsv,Wang:2017htc,Chen:2017ayg,Upadhye:2017hdl,Salvati:2017rsn,
Nunes:2017xon,Emami:2017wqa,Boyle:2017lzt,Zennaro:2017qnp,Sprenger:2018tdb,Wang:2018lun,Giusarma:2018jei,Mishra-Sharma:2018ykh,
Choudhury:2018byy,Choudhury:2018adz,Vagnozzi:2018pwo,Aghanim:2018eyx,Brinckmann:2018owf,Ade:2018sbj,Kreisch:2018var}. The exact figures vary slightly depending on the datasets used, but combinations of some of the most recent and reliable datasets are converging towards a robust 95\%~C.I. upper bound of $M_{\nu}\lesssim 0.15\,\eV$ within the $\lcdm$ model.

It is the goal of this paper to reconsider these bounds if one retreats from the restricted value of $w=-1$ assumed by $\lcdm$. In our work we consider dynamical dark energy with monotonic redshift dependence $w(z)$ given by the standard Chevallier-Polarski-Linder (CPL) parametrization in Eq.~(\ref{cpl})~\cite{Chevallier:2000qy,Linder:2002et}. We use a combination of some of the most recent and robust datasets, which include Cosmic Microwave Background (CMB) measurements from the Planck satellite, Baryon Acoustic Oscillation (BAO) measurements from the SDSS and 6dFGS surveys, and Supernovae Type-Ia (SNeIa) luminosity distance measurements from the JLA catalogue.

One might worry that the neutrino mass bounds could weaken dramatically if the parameter space is enlarged to allow for values of the equation of state other than $w=-1$. Indeed recent work showed that the cosmological bounds on $M_{\nu}$ are weaker when one enlarges the parameter space to other values of $w(z)$ including phantom values $w<-1$. In fact there exists a well-known degeneracy between the DE EoS $w$ and the sum of the three active neutrino masses $M_{\nu}$~\cite{Archidiacono:2013lva,Giusarma:2013pmn,Vagnozzi:2017ovm,Hannestad:2004nb,Lesgourgues:2006nd,Hannestad:2010kz,Wong:2011ip,
Lesgourgues:2012uu,Archidiacono:2017tlz,Lattanzi:2017ubx,Hannestad:2005gj,Goobar:2006xz,Joudaki:2012fx,Lorenz:2017fgo,Lorenz:2017iez,
Sutherland:2018ghu,Sahlen:2018cku}. However, the main result of our paper is that the cosmological bounds on neutrino masses in fact become more restrictive for the case of a DE component with $w(z)\geq-1$ than for the standard $\lcdm$ case of $w=-1$. A comprehensive explanation for this effect will be provided.

From neutrino oscillation data, we know that at least two out of the three neutrino mass eigenstates $m_i$ should be massive, as two different mass splittings between the three active neutrinos are measured. We also know that the smallest mass splitting governs solar neutrino transitions and that it is positive. However, current data are not able to determine the sign of the largest mass splitting, which governs atmospheric neutrino transitions. Therefore, we are left with two possibilities: either the largest mass splitting is positive (normal ordering, NO), or it is negative (inverted ordering, IO). Neutrino oscillation data are currently unable to distinguish among the two possible scenarios. Nevertheless, they impose a lower limit to $\mnu$ of $M_{\nu,\mathrm{min}}\simeq0.1\, {\rm eV}$ within IO and $M_{\nu,\mathrm{min}}\simeq0.06\, {\rm eV}$ within NO~\cite{GonzalezGarcia:2012sz,Gonzalez-Garcia:2014bfa,Bergstrom:2015rba,
Gonzalez-Garcia:2015qrr,Capozzi:2016rtj,Esteban:2016qun,Bilenky:2017rzu,Capozzi:2017ipn,
Caldwell:2017mqu,deSalas:2017kay}.

If one performs a Bayesian analysis, then we find that our bounds imply a mild preference for NO due to parameter space volume effects. Indeed, the available mass range is larger for NO because it goes all the way down to $M_\nu\simeq0.06\,\eV$ rather than only
down to $M_\nu\simeq0.1\,\eV$ for IO. We quantify this preference in terms of probability odds, that we compute following Refs.~\cite{Hannestad:2016fog, Vagnozzi:2017ovm}. If future laboratory experiments determine the mass ordering to be inverted, and if we exclude non-standard physics either in the neutrino or the gravitational sector, one could conclude that the current accelerated expansion of the Universe is likely driven by a component with $w(z)<-1$ within DDE models.

The paper is structured as follows: in Sec.~\ref{sec:dde_par}, we introduce the parametrization adopted for the DDE component, and the conditions imposed on the parameters of the DDE model to satisfy $w(z)\geq-1$; in Sec.~\ref{sec:method}, we present the statistical approach and the dataset employed in this analysis. We discuss the results of this analysis in Sec.~\ref{sec:results} and we finally conclude in Sec.~\ref{sec:conclusion}. For the busy reader who wants to skip to the main results, a summary of the bounds obtained is available in Tab.~\ref{tab:mnu}, and useful visual representations of the same results are also provided in Fig.~\ref{mnu} and Fig.~\ref{wfixed}.

\section{Dynamical dark energy parametrizations}
\label{sec:dde_par}

The simplest parametrization of a DDE component is the Chevallier-Polarski-Linder (CPL) parametrization~\cite{Chevallier:2000qy,Linder:2002et}. In CPL models, the EoS $w$ is parametrized as a function of redshift $z$ as:
\begin{eqnarray}
w_{\text{DDE}}(z) &=& w_0 + w_{\rm a}\frac{z}{1+z} \, ,
\label{cpl}
\end{eqnarray}
where $w_0 = w_{\text{DDE}}(z=0)$ denotes the DE EoS at the present time. This equation corresponds to the first two terms in a Taylor expansion of the EoS in powers of the scale factor $a=1/(1+z)$, around the present time. The truncated expansion of Eq.~(\ref{cpl}) is appropriate if the DE EoS is sufficiently smooth and does not oscillate in cosmic time~\footnote{For models where the DE EoS oscillates, different parametrizations are required, such as those proposed in Refs.~\cite{Lazkoz:2010gz,Ma:2011nc,Pantazis:2016nky} or used in recent observational studies~\cite{Yang:2017alx,Marcondes:2017vjw,Pan:2017zoh}. It is beyond the scope of this work to extend our considerations on the neutrino mass bounds to these types of DE models.}.

It follows from Eq.~(\ref{cpl}) that the non-phantom [$w(z)\geq-1$; NP] condition can be satisfied by imposing the following hard priors:
\begin{eqnarray}
w_0 \geq -1 \, , \quad w_0+w_a \geq -1 \, \,\,\,\, ({\rm NP}).
\label{snpdeprior}
\end{eqnarray}
The first prior imposes the non-phantom condition at present time ($z=0$). The second prior imposes the same condition in the far past, since $\Lim{z \rightarrow \infty}w_{\text{DDE}}(z) = w_0+w_a$. The EoS in Eq.~(\ref{cpl}) is monotonic. Therefore, it is sufficient to impose the NP condition both at the present time and in the far past for the NP condition to hold throughout the Universe expansion history.

The energy density of a  dark energy component corresponding to Eq.~(\ref{cpl}) takes the form:
\begin{eqnarray}
\Omega_{\text{DE}}(z) = \Omega_{\text{DE},0}(1+z)^{3(1+w_0+w_a)}\exp \left ( -3w_a\frac{z}{1+z} \right )\, , \nonumber \\
\label{ocpl}
\end{eqnarray}
where $\Omega_{\text{DE},0}$ is the current dark energy density. The DE component dominates over the other components for $0<z\lesssim z_{m{\rm DE}}$, with $z_{m{\rm DE}} \approx 0.3$ the redshift of matter-dark energy equality. In this range of redshifts, the energy density of a non-phantom dynamical dark energy model with $w(z)\geq-1$ is always greater than that of a corresponding CC model with the same $\Omega_{\text{DE},0}$.

A wide class of smooth non-phantom dynamical dark energy models can be probed if we make use of the EoS given by Eq.~(\ref{cpl}) and we impose the priors in Eq.~(\ref{snpdeprior}). We shall refer to this class of models with the acronym \textbf{\textit{NPDDE}} (\textit{non-phantom dynamical dark energy}). Let us emphasize that the priors in Eq.~(\ref{snpdeprior}) are \textit{crucial} in the derivation of the results we obtain. These priors ensure the stability of the dark energy component and differ from priors considered in previous analyses in the literature~\cite{Zhang:2015uhk,Wang:2016tsz,Zhao:2016ecj,Guo:2017hea,Zhang:2017rbg,Li:2017iur,Yang:2017amu,Peirone:2017lgi,Guo:2018gyo}.

\section{Datasets and Analysis methodology}
\label{sec:method}

We compute constraints on the sum of the three active neutrino masses $M_{\nu}$ with a combination of the most recent cosmological datasets. We consider measurements of the cosmic microwave background (CMB) temperature anisotropies (TT) from the Planck 2015 data release~\cite{Aghanim:2015xee}. We impose a Gaussian prior on the optical depth to reionization of $\tau = 0.055 \pm 0.009$ as a proxy for measurements of CMB polarisation at large scales from the upcoming Planck 2018 release. This prior choice is motivated by the 2016 Planck reanalyses of low-resolution maps in polarization from the Planck High Frequency Instrument~\cite{Aghanim:2016yuo}. In addition to CMB measurements, we consider Baryon Acoustic Oscillation (BAO) measurements from the following catalogues: the SDSS-III BOSS DR11 CMASS and LOWZ galaxy samples~\cite{Anderson:2013zyy}, the DR7 Main Galaxy Sample (MGS)~\cite{Ross:2014qpa}, and the 6dFGS survey~\cite{Beutler:2011hx}. We also include Supernovae Type-Ia (SNeIa) luminosity distance measurements from the SDSS-II/SNLS3 Joint Light-Curve Analysis (JLA) catalogue~\cite{Betoule:2012an,Betoule:2014frx,Mosher:2014gyd}. We refer to the combination of the CMB TT, $\tau$ prior, BAO, and SNeIa datasets as ``\textbf{\textit{base}}''.

We also consider the inclusion of CMB polarization and temperature-polarization spectra (TE,EE) at small scales ($\ell>30$) from the Planck 2015 data release~\cite{Aghanim:2015xee} to the baseline dataset. We refer to the combination of the CMB TT,TE,EE, $\tau$ prior, BAO, and SNeIa datasets as ``\textbf{\textit{pol}}''.

The cosmological model is described by the usual six parameters of the $\Lambda$CDM model: the baryon and cold dark matter physical energy densities $\Omega_{\rm b} h^2$ and $\Omega_{\rm c} h^2$, the angular scale of the acoustic horizon at decoupling $\Theta_{\rm s}$, the optical depth to reionization $\tau$, as well as the amplitude and tilt of the primordial power spectrum $A_{\rm s}$ and $n_{\rm s}$. To this set of parameters, we add the DE EoS parameters $w_0$ and $w_a$, and the sum of the three active neutrino masses $M_{\nu}$. We make use of the publicly available Markov Chain Monte Carlo (MCMC) package \texttt{CosmoMC}~\cite{Lewis:2002ah} to efficiently sample the parameter space.

The treatment of $M_{\nu}$ deserves a further comment. Firstly, we assume three massive degenerate neutrinos, i.e. three massive eigenstates with equal mass $M_{\nu}/3$. This assumption is a valid approximation of the true neutrino mass spectrum, given the current sensitivity of cosmological data~\cite{Gerbino:2016sgw,Archidiacono:2016lnv,Lattanzi:2017ubx}.

Next, we impose a top-hat prior of $M_{\nu}\geq0\, {\rm eV}$. For the purposes of obtaining bounds on neutrino mass from cosmology, we ignore the lower limit $M_{\nu,\mathrm{min}}\simeq0.06\, {\rm eV}$ set by neutrino oscillation experiments~\footnote{We note that the prior $M_{\nu}\geq0\, {\rm eV}$ is in principle improper since it is unconstrained for $M_\nu\rightarrow\infty$ (see~\cite{Tak} for details about proper priors in the astronomy literature). In practice, we adopt a cutoff at $M_{\nu,\mathrm{max}}=3\,\mathrm{eV}$, which makes our prior proper for all intents and purposes. This choice ensures that a large region of the parameter space is sampled, including regions where we already expect the posterior probability to be vanishing from previous experiments (i.e. the region $M_{\nu}>>1\,{\rm eV}$). These unlikely regions of the parameter space are in fact quickly discarded by the MCMC sampling algorithm and only the region of highest posterior probability density is effectively sampled. For the purposes of our analysis, this is equivalent to taking the limit $M_{\nu,\mathrm{max}}\rightarrow\infty$, making our proper prior \textit{de facto} a proxy for the improper prior we describe above. Our result is unaffected by any other choice of sufficiently high value of $M_{\nu\,,\max}$, as long as the posterior probability density for $M_{\nu}>M_{\nu\,,\max}$ is known to be vanishingly small (from previous experiments, analytical considerations, or any other argument).\label{fn:Tak}}. We believe this choice is appropriate for the purpose of this work, because it ensures a bound on $M_{\nu}$ relying exclusively on cosmological data. For recent works discussing different choices of prior on $M_{\nu}$, see for instance Refs.~\cite{Simpson:2017qvj,Schwetz:2017fey,Hannestad:2017ypp,Long:2017dru,Gariazzo:2018pei,Heavens:2018adv,Handley:2018gel,deSalas:2018bym}. 

Moreover, it is fair to say that the only truly a priori information about $M_{\nu}$ is its positivity, i.e. $M_{\nu} \geq 0\,{\rm eV}$. The fact that $M_{\nu} \geq 0.06\,{\rm eV}$ coming from oscillation experiments is not \textit{a priori} but, in fact, \textit{a posteriori} of the oscillation experiments. While the fact that $M_{\nu} \geq 0.06\,{\rm eV}$ can be incorporated as a prior, it is perhaps formally more correct to include it as an external oscillations likelihood~\footnote{We thank the referee for bringing this argument to our attention in a very clear way.}. When viewed from this perspective, it is absolutely clear how our choice of adopting the prior $M_{\nu}\geq0\, {\rm eV}$ is actually independent of the choice of relying exclusively on cosmological data, but rather reflects the only genuine a priori information really present in the problem. The choice of relying exclusively on cosmological data is instead reflected in our choice of not including the oscillations likelihood. Nonetheless, in  Appendix~\ref{app:oscillations}, we briefly discuss the impact including oscillations data. We find that including the oscillations likelihood (in the approximate, but still appropriate to zeroth order, form we choose) has no impact on the conclusions of our work. We further note that for all intents and purposes, as far as upper limits on $M_{\nu}$ (which are the subject of this work) are concerned, the oscillations likelihood can be to zeroth approximation included as a sharp cutoff at $M_{\nu}=0.06\,{\rm eV}$, because the uncertainty to which $M_{\nu,\mathrm{min}}\simeq0.06\, {\rm eV}$ is subject is extremely tiny (but an uncertainty is nonetheless present, whereas the physical lower bound $M_{\nu} \geq 0\,{\rm eV}$ is instead subject o no uncertainty).

In any case, we believe that the approach adopted in this work also allows for a consistency check of the underlying cosmological model. Suppose that we assume a certain cosmological model, and then obtain a cosmological bound on $\mnu$ which lies significantly below $\sim 0.06\, {\rm eV}$. Such a cosmological bound would indicate that either the cosmological model in question is in tension with results from oscillation experiments or that non-standard neutrino physics is required. For example, models with non-standard neutrino interactions leading to a vanishing neutrino energy density today have been proposed~\cite{Beacom:2004yd}. In these cases, the cutoff of the $M_{\nu}$ prior at $M_{\nu}=0\, {\rm eV}$ can be viewed as a phenomenological proxy of the  effect of a lower energy density of neutrinos with respect to the limits imposed by neutrino oscillation measurements. Note that we are implicitly excluding the possibility that such a finding could be a signal for unaccounted systematics in the dataset employed.

Finally, we combine results from cosmology and neutrino oscillation experiments to quantify the preference for one of the two neutrino mass orderings. In this part of the work, we do impose lower bounds on neutrino mass from oscillation experiments, although we previously did not use them in obtaining bounds on neutrino masses.  We follow the Bayesian approach illustrated in~\cite{Hannestad:2016fog,Vagnozzi:2017ovm}. We denote by $\pi(I),\,\pi(N)$ the \textit{prior probabilities} for the normal and inverted ordering respectively. Then, we compute $p_O$ (where $O=N,\,I$), the \textit{posterior probabilities} for each of the two orderings, as follows:
\begin{widetext}
\begin{eqnarray}
\hskip -0.6 cm p_O = \frac{\pi(O)\int_0^{\infty} dm_0 \ {\cal L}(D \vert m_0, O)}{\pi(N)\int_0^{\infty} dm_0 \ {\cal L}(D \vert m_0, N)+\pi(I)\int_0^{\infty} dm_0 \ {\cal L}(D \vert m_0, I)} \, .
\label{eq:evidence}
\end{eqnarray}
\end{widetext}
In Eq.~(\ref{eq:evidence}), $m_0$ is the mass of the lightest neutrino eigenstate and ${\cal L}(D \vert m_0, O)$ is the likelihood of cosmological data $D$. The above Eq.~(\ref{eq:evidence}) implicitly assumes a cutoff $m_{0\,,\max}\rightarrow\infty$ in the prior probability for $m_0$. Moreover, $m_{0\,,\max}$ should take the same value for both orderings and be large enough so as not to cut the prior in a region where the posterior would otherwise be significantly different from zero (so that it is effectively the data through the likelihood, rather than the prior itself, which cuts the region of high $m_0$, see Footnote~(\ref{fn:Tak}) for a previous related discussion when considering the formally improper prior on $M_{\nu}$). In this way, both the numerator and the denominator should formally contain a $1/m_{0\,,\max}$ normalization factor, which then cancels out when taking the ratio appearing in Eq.~(\ref{eq:evidence}). We take ${\cal L}(D \vert m_0, O)$ from the analysis of cosmological data illustrated in this work. We use the values of $\pi(I),\,\pi(N)$ from the global Bayesian analysis of neutrino oscillation measurements~\cite{Bergstrom:2015rba}. For further details about how to compute ${\cal L}(D \vert m_0, O)$ and get to Eq.~(\ref{eq:evidence}), we refer the reader to the thorough discussions in~\cite{Hannestad:2016fog, Vagnozzi:2017ovm}. We convey the results in terms of probability odds of normal versus inverted ordering ($p_{\rm NO}:p_{\rm IO}$).

We follow the approach of~\cite{Hannestad:2016fog,Vagnozzi:2017ovm} as it is a quick, yet reliable, way to quantify the preference for the normal ordering in different cosmological scenarios. The method used in this work should be kept in mind when one compares the results quoted here with results from other works. Indeed, we remind that alternative approaches can be adopted to quantify the statistical preference for the neutrino mass ordering~\cite{Gerbino:2016ehw,Simpson:2017qvj,Capozzi:2017ipn}. For the sake of comparison, in Appendix~\ref{app:profile} we report an alternative estimate of the sensitivity to the mass ordering based on the Akaike information criterion (AIC). The specific outcomes of each analysis should be interpreted only in light of the method adopted.

\section{Results}
\label{sec:results}
In this section, we present the bounds on the sum of the three active neutrino masses; we provide a thorough physical explanation of the results; we discuss the Bayesian statistical approach we have used, as well as the dependence of our results on this approach; and we conclude by commenting on the implications of our results for the determination of the neutrino mass ordering.

\subsection{Bounds on Neutrino Masses}
\label{subsec:bounds}

Table~\ref{tab:mnu} shows the bounds on the sum of the neutrino masses $M_{\nu}$ for three cases: a) a dark energy component satisfying the dominant energy condition, with equation of state (EoS) $w(z)\geq-1$ throughout the expansion history of the Universe (non-phantom dynamical dark energy, NPDDE); b) the standard cosmological model ($\lcdm$) with cold dark matter and a cosmological constant where $w(z)=-1$ is fixed; c) a generic dynamical dark energy (DDE) model with EoS given by the CPL parametrization, Eq.~(\ref{cpl}), with $w_0$ and $w_a$ free to vary even within the phantom region where $w(z)<-1$. We refer to this last model as $w_0w_a$CDM. Constraints on $\mnu$ are presented for the two different combinations of cosmological datasets, \textit{base} and \textit{pol}, described at the beginning of Sec.~\ref{sec:method}.

\begin{table*}
\begin{tabular}{|l|c|c|c|}
\hline
	&$w(z)\geq-1,\,(\mathrm{NPDDE})$	&$w(z)=-1,\,(\lcdm)$	&$w_0w_a$CDM\\
\hline
$\mathrm{dataset:}\,\textit{base}$	&$\mnu<0.13\,\eV$	&$\mnu<0.16\,\eV$	&$\mnu<0.41\,\eV$\\
$\mathrm{dataset:}\,\textit{pol}$	&$\mnu<0.11\,\eV$	&$\mnu<0.13\,\eV$	&$\mnu<0.37\,\eV$\\
\hline
\end{tabular}
\caption{95\%~C.I. upper bounds on the sum of the neutrino masses $\mnu$. Columns correspond to the different cosmological models assumed in this work: a) a dark energy component satisfying the dominant energy condition, with equation of state (EoS) parametrized through Eq.~(\ref{cpl}) and satisfying $w(z)\geq-1$ throughout the expansion history of the Universe (non-phantom dynamical dark energy, NPDDE); b) the standard cosmological model ($\lcdm$) with cold dark matter and a cosmological constant where $w(z)=-1$ is fixed; c) a generic dynamical dark energy (DDE) model with EoS given by the CPL parametrization, Eq.~(\ref{cpl}), with $w_0$ and $w_a$ free to vary even within the phantom region where $w(z)<-1$ ($w_0w_a$CDM). Rows report the constraints on $\mnu$ for two different combinations of cosmological datasets, \textit{base} and \textit{pol}. These two combinations include CMB, BAO and SNeIa data, and they only differ in the use of CMB polarization data at small scales, as described in the text at the beginning of Sec.~\ref{sec:method}.}
\label{tab:mnu}
\end{table*}

For the $\Lambda$CDM model, we find 95\%~C.I. upper bounds of $M_{\nu}<0.16\,\eV$ for the \textit{base} dataset and $\mnu<0.13\,\eV$ for the \textit{pol} dataset. When we instead assume the more generic $w_0w_a$CDM model, that also allows for $w(z)<-1$, the 95\%~C.I. upper bound on $M_{\nu}$ is significantly relaxed to $\mnu<0.41\,\eV$ for the \textit{base} dataset and $M_{\nu}<0.37\,{\rm eV}$ for the \textit{pol} dataset. These broader bounds are expected, given the well known degeneracy between $M_\nu$ and an arbitrary DDE component.

We now consider a NPDDE model and impose $w(z)\geq-1$ throughout the expansion history. In this case, we find the stringent 95\%~C.I. upper bounds of $\mnu<0.13\,\eV$ for the \textit{base} dataset and $M_{\nu}<0.11\, {\rm eV}$ for the \textit{pol} dataset. Therefore, we find that \textbf{\textit{the constraints on the sum of the neutrino masses in dynamical dark energy models with $w(z)\geq-1$ are slightly tighter than those obtained in}} $\boldsymbol{\Lambda{\rm CDM}}$, despite the enlarged parameter space (two extra parameters) in NPDDE models. We note that the upper bounds found within the NPDDE model are also very close to the minimal mass allowed in the inverted ordering scenario, $M_{\nu,\mathrm{min}}\simeq0.1\,\eV$.

\begin{figure}[!h]
\includegraphics[width=0.9\linewidth]{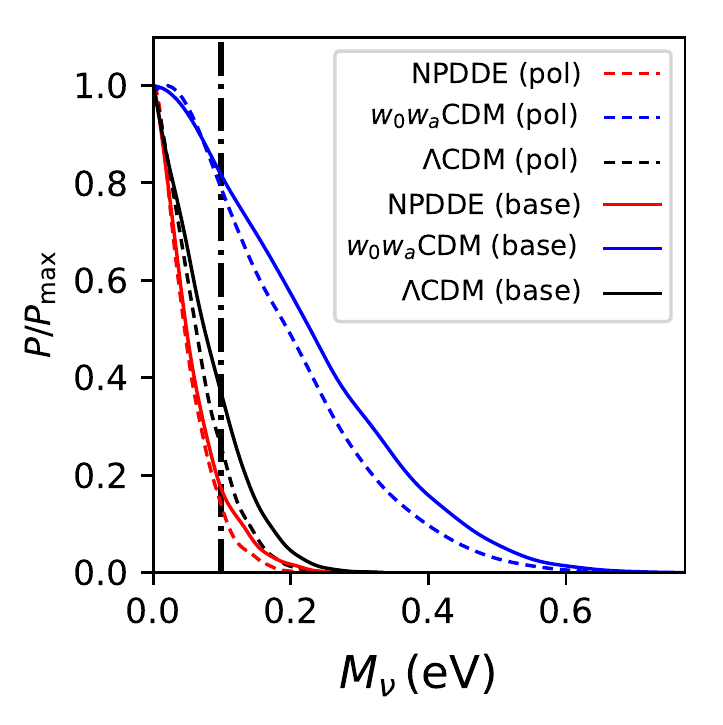}
\caption{One-dimensional posterior probabilities of the sum of the three active neutrino masses $M_{\nu}$ (in eV)  for three cases:  the $w_0w_a$CDM generic DDE case which allows for values of $w$ both smaller than or larger than $-1$ (in blue), the $\lcdm$ case (in black), and the non-phantom dynamical dark energy (NPDDE) model with $w(z)\geq-1$ (in red). Results have been obtained using a Bayesian analysis that marginalizes over all applicable $w_0, w_a$  values, and are shown for the two dataset combinations employed in this work as described at the beginning of Sec.~\ref{sec:method}: solid for \textit{base} (using CMB, BAO, and SN data), dashed for \textit{pol} (also including CMB polarization at small scales). The vertical black dotted-dashed line corresponds to the minimal mass of $M_{\nu,\min} \approx 0.1\, {\rm eV}$ allowed by neutrino oscillation data within the inverted ordering.}

\label{mnu}
\end{figure}

Figure~\ref{mnu} depicts the one-dimensional posterior probabilities of $M_{\nu}$ for the $w_0w_a$CDM generic DDE case (in blue), the $\lcdm$ case (in black), and the NPDDE model with $w(z)\geq-1$ (in red). Results for the two dataset combinations employed in this work are shown: solid for \textit{base}, dashed for \textit{pol}. For each dataset combination, the significant shift of the upper bounds on $M_{\nu}$ to smaller values is visually clear as one moves from the blue to the red curves. The vertical black dotted-dashed line corresponds to the minimal mass of $M_{\nu,\min} \approx 0.1\, {\rm eV}$ allowed by neutrino oscillation data within the inverted ordering.

\subsection{Physical Explanation of Results}
\label{subsec:physics}

We have observed that the bounds on $\mnu$ in the NPDDE model are not weaker -- and actually slightly tighter -- than those in $\lcdm$. Here we provide the physical explanation for these results. The reader can refer to Refs.~\cite{Hannestad:2004nb,Lesgourgues:2006nd,Hannestad:2010kz,Wong:2011ip,Lesgourgues:2012uu,Archidiacono:2017tlz,
Lattanzi:2017ubx} for comprehensive reviews of the effects of massive neutrinos in cosmology~\footnote{Modifying the assumed expansionary history of the Universe will generically lead to different conclusions concerning $M_{\nu}$. The reason is that the effect of $M_{\nu}$ on cosmological data is degenerate with other parameters governing the expansionary history, such as the DE EoS $w$: in other words, the effect on cosmological observables of a change in $M_{\nu}$ (for instance, the resulting change in the distance to last-scattering, further discussed later in this Section) can be compensated by adjusting these other parameters. Therefore, an expansionary history which is different from $\Lambda$CDM leads to bounds on $M_{\nu}$ which are different from those obtained assuming $\Lambda$CDM. Obviously, the same is true if the expansionary history is restricted to a class of models of which $\Lambda$CDM represents a particular case: in our case, $\Lambda$CDM represents the particular case of the NPDDE class of models, when $w_0=-1$ and $w_a=0$.}.

The CMB temperature data accurately constrain the position and amplitude of the first acoustic peak in the CMB power spectrum. These constraints entail a very precise determination of the angular size of the sound horizon at decoupling $\Theta_{\rm s}$ and of the redshift of matter-radiation equality $z_\mathrm{eq}$. Therefore, any change in the DE sector should be compensated by shifts in the other cosmological parameters such that $\Theta_{\rm s}$ and $z_\mathrm{eq}$ remain approximately fixed.

The angular size of the first peak $\Theta_{\rm s}$ is defined as the ratio between the sound horizon at decoupling $r_s$ and the angular diameter distance to last scattering $D_A$. The sound horizon at decoupling $r_s$ is essentially fixed by pre-recombination physics. It is thus unaffected by changes in the dark energy sector, which are only relevant at late times. The angular diameter distance to last scattering $D_A$ is instead sensitive to the late-time evolution of the Universe. Therefore, $D_A$ is affected by the physics of dark energy.

In order to keep $\Theta_{\rm s}$ unchanged in the NPDDE framework, it is necessary that $D_A$ remains fixed as well. Up to proportionality factors, $D_A$ is given by:
\begin{eqnarray}
D_A(z_{\text{LS}}) \propto \frac{1}{H_0}\int_{0}^{z_{\text{LS}}} \frac{dz}{E(z)} \, ,
\label{da}
\end{eqnarray}
where $z_{\text{LS}}$ denotes the redshift of last-scattering. The function $E(z)$ denotes the Hubble parameter at redshift $z$ normalized by its value today. In the NPDDE model it is given by:
\begin{eqnarray}
E(z) &=& \frac{H(z)}{H_0} \nonumber \\
& \simeq & \sqrt{(\Omega_c+\Omega_b)(1+z)^3+\Omega_{\text{DE}}(z)+\Omega_{\nu}(z)} \, .
\label{eq:normH}
\end{eqnarray}
In the above equation, $\Omega_c$ and $\Omega_b$ are the current cold dark matter and baryon energy densities respectively, $\Omega_{\text{DE}}(z)$ is the dark energy density given by Eq.~(\ref{ocpl}), and $\Omega_{\nu}(z)$ is the neutrino energy density. At late times, after neutrinos become nonrelativistic, $\Omega_{\nu}h^2 \approx M_{\nu} (1+z)^3 /93.14\,{\rm eV}$, where $h=H_0/(100\,{\rm km}\,{\rm s}^{-1}\,{\rm Mpc}^{-1})$. In writing Eq.~(\ref{eq:normH}), we are neglecting the contribution of the photon energy density $\Omega_\gamma$, which is negligible at the redshifts under consideration.

It is easy to show that \textit{the normalized expansion rate $E(z)$ at late times is higher in a NPDDE Universe than in a $\Lambda$CDM one} for fixed values of $\Omega_c$, $\Omega_b$, $M_{\nu}$ and $\Omega_{\text{DE},0}$, as well shall comment more on below and in Fig.~\ref{e}. The integral in Eq.~(\ref{da}) at fixed $M_{\nu}$ is therefore smaller in the NPDDE case than in the $\lcdm$ one. In order to keep $D_A$ fixed, one is left with the option of decreasing both $H_0$ and $M_{\nu}$. This option is preferred over the choice where one parameter is decreased by a greater amount while the other parameter is kept fixed, since in the latter case the more sizeable decrease of the first parameter can lead to undesired changes in other regions of the CMB spectra, despite $\Theta_{\rm s}$ being kept fixed. One could argue that the same effect can be obtained by decreasing $\Omega_c$ and/or $\Omega_b$. However, this choice would alter the redshift of matter-radiation equality, which is accurately constrained by the amplitude of the first acoustic peak in the CMB power spectrum. Therefore, it is not the preferred choice. This physical explanation for the shifts in the bounds of $M_{\nu}$ is fully supported by the results of our Monte Carlo analyses. In particular, we have verified that the posterior of $\Theta_{\rm s}$ is nearly unchanged when moving from the $\Lambda$CDM scenario to the NPDDE model.


From the explanation above it follows that in NPDDE models we expect a lower Hubble constant $H_0$ and/or a lower sum of the neutrino masses $M_{\nu}$ compared to the $\lcdm$ case. The shifts in $H_0$ and $\mnu$ are necessary to keep $\Theta_{\rm s}$ fixed. Therefore, the very strong anti-correlation (degeneracy) between $M_{\nu}$ and $H_0$ present in $\Lambda$CDM is weakened in NPDDE models.  Note also that in NPDDE models we expect a lower value of $\sigma_8$, thus reducing the tension between primary CMB and cluster counts/weak lensing measurements (see e.g.~\cite{Heymans:2012gg,Kilbinger:2012qz,Heymans:2013fya,Macaulay:2013swa,MacCrann:2014wfa,Raveri:2015maa,Joudaki:2016mvz,
Kitching:2016hvn,Hildebrandt:2016iqg,Ko:2016uft,Leauthaud:2016jdb,Camera:2017tws,DiValentino:2017oaw,Gomez-Valent:2017idt,Gomez-Valent:2018nib} for some works examining this tension and possible solutions).

\begin{figure}[!h]
\includegraphics[width=0.9\linewidth]{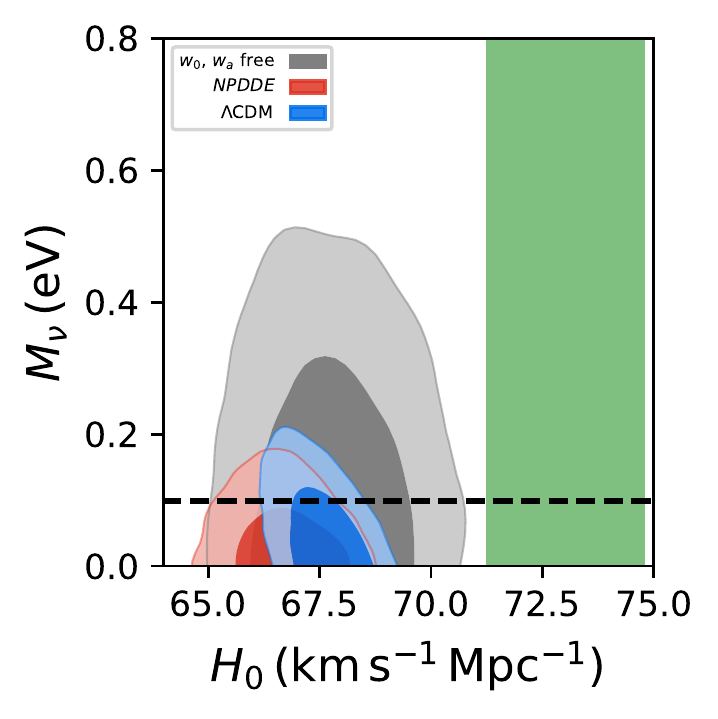}
\caption{Two-dimensional probability contours in the $H_0-M_{\nu}$ plane. The blue contours are obtained for the $\Lambda$CDM model, the red contours are for a dynamical dark energy model with EoS parametrized by Eq.~(\ref{cpl}) and satisfying $w(z)\geq-1$ (NPDDE), and the grey contours are for a generic dark energy model with EoS parametrized by Eq.~(\ref{cpl}). The green band indicates the 68\%~C.I. for $H_0$ from direct measurements of the Hubble Space Telescope~\cite{Riess:2016jrr,Riess:2011yx}. The horizontal dashed line corresponds to $M_{\nu,\mathrm{min}}\simeq0.1\,\eV$, the minimal value for the sum of the neutrino masses allowed in the inverted ordering scenario by neutrino oscillation data. When moving from the $\Lambda$CDM contours (blue) to models with $w(z)\geq-1$ (red), the shifts of $H_0$ and $M_{\nu}$ to smaller values are evident. These shifts are necessary to keep the angular scale of the sound horizon at recombination $\Theta_{\rm s}$ fixed, see discussion in the main text. It is also clear that the $M_{\nu}$-$H_0$ degeneracy is weakened when moving from $\Lambda$CDM to models with $w(z)\geq-1$ (NPDDE). For further information, see discussion in main text concerning the $M_{\nu}$-$H_0$ correlation coefficient, which is reduced from $-0.43$ ($\Lambda$CDM) to $-0.14$ (NPDDE). The tension between direct measurements of $H_0$ and cosmological estimates is not resolved by a dark energy component with $w(z)\geq-1$. The tension is partially alleviated by a generic dark energy component which can access the $w(z)<-1$ region (grey contours). The contour regions are obtained for the \textit{base} dataset combination of CMB, BAO and SNeIa data, with no CMB small scale polarization data. Similar considerations apply to the contours derived from the combination which also includes small scale CMB polarization data.}
\label{2D_mnu_H0}
\end{figure}

\begin{figure}[!h]
\includegraphics[width=0.9\linewidth]{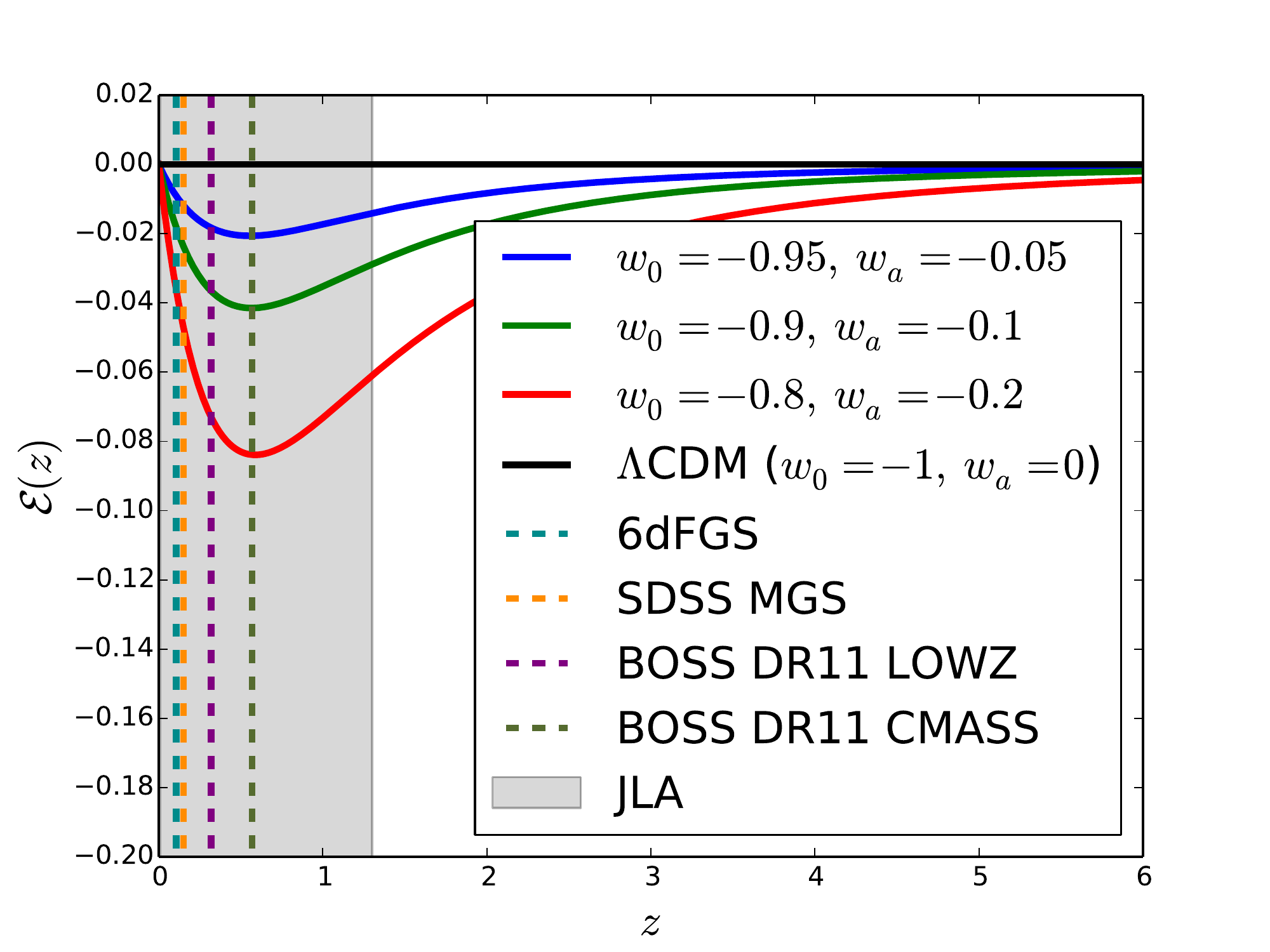}
\caption{${\cal E}(z)$, defined in Eq.~(\ref{ez}), quantifies the difference in the normalized expansion rate $H(z)/H_0$ between a dynamical dark energy model with equation of state $w(z)\geq-1$ (NPDDE) and a $\Lambda$CDM model.  The quantity ${\cal E}(z)$ is plotted for sample cosmologies with $w_0=-0.95,\,w_a=-0.05$ (blue curve), $w_0=-0.9,\,w_a=-0.1$ (green curve),  $w_0=-0.8,\,w_a=-0.2$ (red curve), and $w_0=-1,\,w_a=0$ (black curve, $\Lambda$CDM, where ${\cal E}(z)=0$). We have fixed $\Omega_{m,0}=0.3$ and $\Omega_{\text{DE},0}=0.7$. The negative ${\cal E}(z)$ indicates that the normalized expansion rate is higher in the NPDDE model compared to $\Lambda$CDM. The four vertical dashed lines indicate the redshift of the four BAO measurements we consider in this work: 6dFGS (cyan), SDSS MGS (orange), BOSS DR11 LOWZ (purple), and BOSS DR11 CMASS (green). The grey shaded band refers to the redshift coverage of the JLA Supernovae Ia sample. Thus the measurements considered in this work probe the redshift range in which the dip in ${\cal E}(z)$ is most prominent.}
\label{e}
\end{figure}

In Fig.~\ref{2D_mnu_H0} we show the two-dimensional joint $H_0$-$M_{\nu}$ posterior for the \textit{base} dataset. The blue contours are obtained in the $\lcdm$ model, the red contours in the NPDDE model where $w(z)\geq-1$, and the grey contours for the more generic $w_0w_a$CDM model where also $w(z)<-1$ is allowed. The horizontal dashed line corresponds to $M_{\nu,\mathrm{min}}\simeq0.1\,\eV$, the minimal value allowed by neutrino oscillation data in the inverted ordering scenario. The difference between the blue contours ($\lcdm$) and the red contours (NPDDE model) is compatible with the shifts in $H_0$ and $M_{\nu}$ required to keep $\Theta_{\rm s}$ fixed. The green band in Fig.~\ref{2D_mnu_H0} corresponds to the 68\%~C.I. for $H_0$ inferred by direct measurements from the Hubble Space Telescope~\cite{Riess:2011yx,Riess:2016jrr}. From Fig.~\ref{2D_mnu_H0}, it is clear that the tension between direct measurements and cosmological estimates of $H_0$ is not resolved, and actually worsened, within a NPDDE model. The tension can be partially alleviated by a generic dark energy component (the $w_0w_a$CDM model) able to access the region $w(z)<-1$ (grey contours)~\cite{DiValentino:2015ola,DiValentino:2016hlg,Bernal:2016gxb,Zhao:2017cud,DiValentino:2017gzb,
Salvatelli:2013wra,Qing-Guo:2016ykt,Karwal:2016vyq,Kumar:2016zpg,Shafieloo:2016bpk,Kumar:2017dnp,
Feng:2017nss,Zhao:2017urm,DiValentino:2017zyq,DiValentino:2017iww,Yang:2017yme,Feng:2017mfs,Yang:2017ccc,
Wang:2017lai,DiValentino:2017rcr,Pan:2017ent,Feng:2017usu,Matsumoto:2017qil,Wang:2018ahw,Mortsell:2018mfj,Poulin:2018zxs,
Yang:2018ubt,Yang:2018euj,DiValentino:2018wum,Yang:2018xlt}~\footnote{A notable exception to this statement is however provided by running vacuum models (RVM), motivated by quantum field theoretical considerations~\cite{Sola:2013gha}. Studies have shown that RVMs, which appear to be statistically preferred over $\Lambda$CDM, can address the $H_0$ tension, but do so invoking a non-phantom dark energy component (with the value $w=-1$ being preferred, see e.g.~\cite{Sola:2015wwa,Sola:2016jky,Sola:2016ecz,Sola:2016zeg,Sola:2016hnq,Sola:2017jbl,Sola:2017znb,Sola:2017lxc,Gomez-Valent:2017idt,Gomez-Valent:2018nib} for some recent works). Notice that RVMs point towards values of $H_0$ which are closer to the CMB inferred value.}. We have checked that similar considerations apply to the corresponding contour plot obtained with the \textit{pol} dataset.

From Fig.~\ref{2D_mnu_H0} we also see that the anti-correlation (degeneracy) between $M_{\nu}$ and $H_0$ is weakened when moving from $\Lambda$CDM (blue) to NPDDE models (red). The magnitude of the degeneracy is reflected by the tilt of the main axes of the ellipsoidal $M_{\nu}$-$H_0$ contours. The contour in the $\Lambda$CDM case is visibly more inclined than the NPDDE one. The weakening of the $M_{\nu}$-$H_0$ degeneracy can be rigorously quantified by computing the correlation coefficient between the two parameters. The correlation coefficient between two parameters $i$ and $j$, $R_{ij}$, is defined as $R_{ij} = C_{ij}/\sqrt{C_{ii}C_{jj}}$, where $C$ is the covariance matrix of the cosmological parameters (in our case $i=M_{\nu}$ and $j=H_0$), estimated from our MCMC runs. For the \textit{base} [\textit{pol}] dataset, we find a correlation coefficient of $-0.43$ [$-0.50$] in the $\Lambda$CDM case, which is lowered to $-0.14$ [$-0.16$] in the NPDDE case. Therefore, the correlation between the two parameters is strongly reduced in moving from $\Lambda$CDM to NPDDE models.

We shall now demonstrate that the late-time expansion rate $E(z)$ is higher in a Universe with $w(z)\geq-1$ compared to $\lcdm$. We shall also identify the redshift range in which this effect is most prominent. We define the following quantity:
%
%
%
\begin{eqnarray}
{\cal E}(z) \equiv \left ( \frac{E(z)\vert_{\Lambda{\rm CDM}}}{E(z)\vert_{\rm NPDDE}} \right )^2\Bigg\vert_{\Omega_m\,,\,\Omega_{\text{DE},0}} - 1 \, ,
\label{ez}
\end{eqnarray}
where $\vert_{\Lambda{\rm CDM}}$ and $\vert_{\rm NPDDE}$ indicate that $E(z)$ is evaluated in a $\Lambda$CDM Universe or in a Universe with $w(z)\geq-1$ respectively. The notation $\vert_{\Omega_m\,,\,\Omega_{\text{DE},0}}$ denotes that $\Omega_m=\Omega_c+\Omega_b+\Omega_{\nu}$ and $\Omega_{\text{DE},0}$ are kept fixed when moving from $\Lambda$CDM to NPDDE. ${\cal E}(z)=0$ therefore corresponds to the $\Lambda$CDM case. A negative ${\cal E}(z)$ instead indicates that the expansion rate normalized by $H_0$ is higher in the NPDDE model compared to $\Lambda$CDM. Note that ${\cal E}(z)$ is closely related to other diagnostics used in the literature to probe the DE evolution, such as the $Om$ diagnostic~\cite{Sahni:2008xx}. In Fig.~\ref{e}, ${\cal E}(z)$ is plotted for three choices of $w_0,\,w_a$. All of the choices satisfy the stability priors imposed by Eq.~(\ref{snpdeprior}) and ensure that $w(z)\geq-1$.

Figure~\ref{e} clearly shows that ${\cal E}(z)$ is negative at low redshifts, as expected from the above discussion. $\mathcal{E}(z)$ also shows a minimum for $z \approx 0.5$ for values of $w_0$ and $w_a$ that are allowed by cosmological data. The four vertical dashed lines indicate the redshift of the four BAO measurements we consider in this work. The grey shaded band refers to the redshift coverage of the JLA Supernovae Type-Ia sample we consider in this analysis. Thus, we see that the measurements adopted in this work cover the redshift range where the difference between ${\cal E}(z)$ and ${\cal E}(z)=0$ is largest. Therefore, the redshift range of current BAO and SNeIa measurements is ideal to probe the dynamics of non-phantom [$w(z)\geq-1$] dark energy.

\subsection{Comment on the Bayesian statistical approach adopted}
\label{subsec:comment}

Here we comment on the a priori counterintuitive fact that the bounds on $M_{\nu}$ in the NPDDE model are tighter than those in $\Lambda$CDM, despite the fact that $\Lambda$CDM represents the limiting case of NPDDE when $w_0=-1$ and $w_a=0$. Indeed  these tighter bounds are a result of our use of a Bayesian statistical approach~\cite{Trotta:2008qt,Verde:2009tu}.

To explain our results, we begin by \textit{fixing} the parameters $w_0$ and $w_a$ to specific values not corresponding to $\lcdm$ (i.e., $w_0 \neq -1$ and $w_a \neq 0$), yet still satisfying $w(z)\geq-1$.  Following the explanation of the previous Section, we expect that the bounds on $M_\nu$ must become ever tighter as the dark energy model gets farther away from $\Lambda$CDM.  We will study specific cases below and find that these expectations are met. Therefore, a Bayesian analysis marginalizing over the range of $w_0$, $w_a$ values satisfying $w(z)\geq-1$ is expected to obtain a bound on $M_{\nu}$ which is slightly tighter than the $\lcdm$ one, as shown by the results in Sec.~\ref{subsec:bounds}

Specifically we considered four test cases: a) $w_0=-0.95$, $w_a=0$, b) $w_0=-0.95$, $w_a=0.05$, c) $w_0=-0.9$, $w_a=0$, and d) $w_0=-0.85$, $w_a=0$, and found 95\%~C.I. upper bounds of a) $M_\nu<0.13\,\eV$, b) $M_\nu<0.12\,\eV$, c) $M_\nu<0.11\,\eV$ and d) $M_\nu<0.08\,\eV$. Indeed the bounds on neutrino mass are tighter than in the case of standard $\Lambda$CDM.
 
\begin{figure}[!h]
\includegraphics[width=0.9\linewidth]{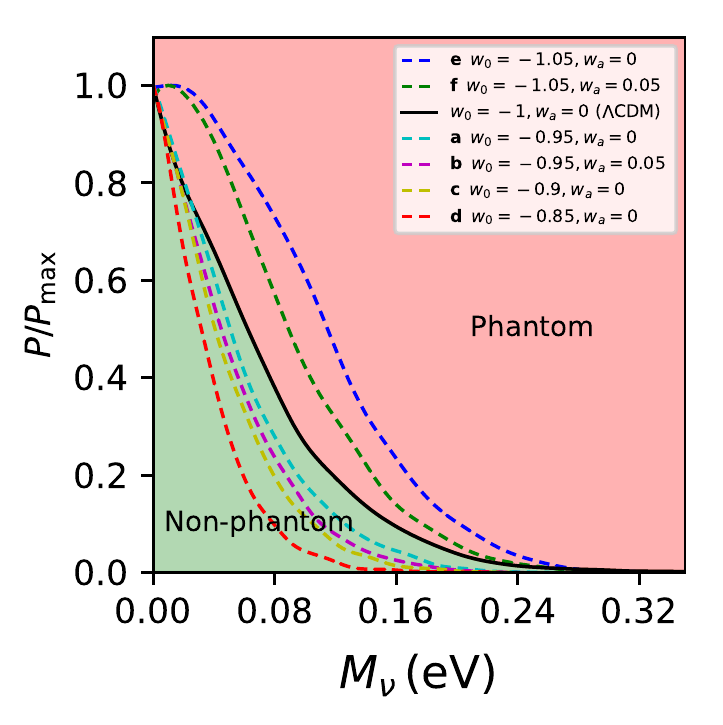}
\caption{One-dimensional posterior probabilities of the sum of the three active neutrino masses $M_{\nu}$ (in eV) for a selection of cosmological models with $w_0$ and $w_a$ \textit{fixed}, described in Sec.~\ref{subsec:comment}. Models a)-d) have $w_0$ and $w_a$ fixed to values satisfying the condition $w(z)\geq-1$, and are represented by the dashed light blue, dashed purple, dashed yellow, and dashed red curves respectively. Models e) and f) have $w_0$ and $w_a$ fixed to values not satisfying the condition $w(z)\geq-1$, and are represented by the dashed dark blue and dashed green curves respectively. The $\lcdm$ result corresponds to the solid black line. The region where $w(z)\geq-1$ is satisfied is shaded in green and labeled ``Non-phantom''; conversely, the region where $w(z)\geq-1$ is not satisfied is shaded in pink and labeled ``Phantom''. It is clear that the bounds on $M_{\nu}$ for models where $w_0$ and $w_a$ are fixed to values satisfying $w(z)\geq-1$ are always tighter than the $\lcdm$ bound. Therefore, a Bayesian analysis marginalizing over the range of $w_0$, $w_a$ values satisfying $w(z)\geq-1$ is expected to obtain a bound on $M_{\nu}$ which is slightly tighter than the $\lcdm$ one, as shown by the results in Sec.~\ref{subsec:bounds}.}
\label{wfixed}
\end{figure}

\begin{figure}[!h]
\includegraphics[width=0.9\linewidth]{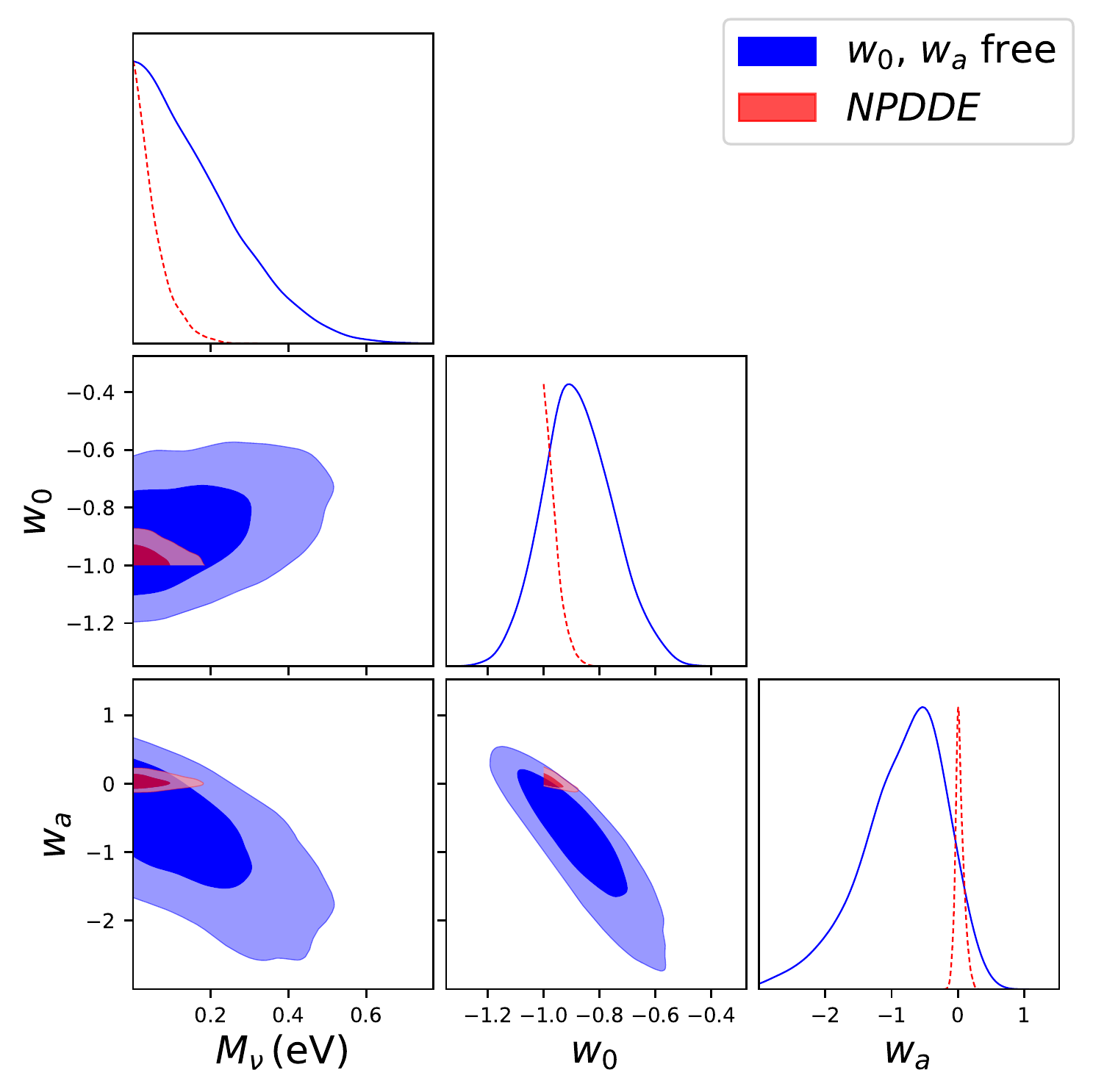}
\caption{$68\%$~C.L. (dark blue/red) and $95\%$~C.L. (light blue/red) joint posterior distributions in the $M_{\nu}$-$w_0$-$w_a$ plane, along with their marginalized posterior distributions from the \textit{base} dataset, for the $w_0w_a$CDM (blue contours) and NPDDE models (red contours). The marginalized posterior distributions appearing along the diagonal are normalizable probability distributions and hence in arbitrary units. The sharp cuts in the red posteriors are due to the hard NPDDE priors [see Eq.~\ref{snpdeprior}].}
\label{cp}
\end{figure}

The posterior distributions of $M_{\nu}$ are shown in Fig.~\ref{wfixed}, in dashed light blue, dashed purple, dashed yellow, and dashed red for cases a)-d) respectively. The $\lcdm$ bound is instead represented by the solid black line. It is visually clear that the bounds for these cases are all tighter than the $\lcdm$ one. For pedagogical purposes we have also considered two cases where $w_0 \neq -1$ and $w_a \neq 0$ are instead fixed to values such that $w(z)\geq-1$ is not satisfied: e) $w_0=-1.05$, $w_a=0$, and f) $w_0=-1.05$, $w_a=0.05$. The corresponding posterior distributions are shown in dashed dark blue and dashed green in Fig.~\ref{wfixed} and corresponds to 95\%~C.I. upper bounds of e) $M_\nu<0.19\,\eV$ and f) $M_\nu<0.18\,\eV$. As per our expectations, the bounds for cases e) and f) are looser than the $\lcdm$ one. To further back up this argument, we show a triangular plot in $M_{\nu}$-$w_0$-$w_a$ space in Fig.~\ref{cp}, where we compare constraints obtained assuming the $w_0w_a$CDM model (blue contours) and the NPDDE model (red contours). From there it is clear that restricting the allowed region to the NPDDE parameter space inevitably selects the region of parameter space with very low $M_{\nu}$, due to the direction of the mutual $M_{\nu}$-$w_0$-$w_a$ degeneracies.

In the Bayesian statistical approach adopted to obtain the results in Sec.~\ref{subsec:bounds}, $w_0$ and $w_a$ are not fixed, but rather varied. Subsequently, the uncertainty in $w_0$ and $w_a$ is integrated out by the process of marginalization, leading to the marginalized posterior on $M_{\nu}$. Heuristically, this procedure can be viewed as a weighted average over the range of prior possibilities of $w_0$ and $w_a$, with weights given by the value of the prior in that particular point of parameter space. For each of these prior possibilities of $w_0$ and $w_a$, we have already seen that the corresponding bound on $M_{\nu}$ is tighter than the $\lcdm$ bound, see examples a)-d) above as well as the green-shaded region in Fig.~\ref{wfixed}. Therefore, the weighted average of such bounds is expected to be not weaker than the $\lcdm$ constraint, as confirmed by our results in Sec.~\ref{subsec:bounds}. This explains the a priori counter-intuitive fact that the upper limits on $M_{\nu}$ for the NPDDE model are slightly tighter than the $\lcdm$ limit despite the enlarged parameter space.

\subsection{Implications for the determination of the neutrino mass ordering}
\label{subsec:determination}

Finally, we comment on the implications that the results of this work could have for the determination of the neutrino mass ordering. By integrating the posterior distributions of $\mnu$ for both the \textit{base} and \textit{pol} datasets (solid red and dashed red lines in Fig.~\ref{mnu} respectively) it is straightforward to show that a significant fraction ($\gtrsim 90\%$) of the $M_{\nu}$ posterior probability lies in the range $M_{\nu}<0.1\, {\rm eV}$. This region is precluded to the inverted mass ordering by neutrino oscillation data.

Should non-cosmological probes such as long-baseline neutrino oscillations experiments (for example T2K~\cite{Abe:2011ks}, NOvA~\cite{Patterson:2012zs}, or DUNE~\cite{Acciarri:2015uup}) establish that the neutrino mass ordering is inverted, the viability of dark energy models with $w(z)\geq-1$ could be jeopardized. This conclusion holds if we exclude exotic physics at play in the cosmological neutrino sector and/or in the gravitational sector. Examples of such exotic models are those with non-standard neutrino interactions predicting a vanishing neutrino energy density today~\cite{Beacom:2004yd} or mass-varying neutrinos~\cite{Fardon:2003eh,Franca:2009xp,Wetterich:2013jsa,Geng:2015haa}, and models of modified gravity where the bounds on $M_{\nu}$ could be significantly different from those in $\lcdm$~\cite{Knox:2005rg,Huterer:2006mva,Kunz:2006ca,Hu:2014sea,Battye:2015hza,Bellomo:2016xhl,
Dirian:2017pwp,Renk:2017rzu,Peirone:2017vcq}.

Therefore, we have brought to light a subtle and perhaps unexpected connection between two at first glance seemingly disconnected fields: neutrino oscillation experiments and the nature of dark energy. In the near future, results from the former might be able to shed important light on the latter. It is also worth noticing that our findings could also be very interesting in light of the recently revived Swampland conjectures~\cite{Vafa:2005ui,Obied:2018sgi,Agrawal:2018own} (see also e.g.~\cite{Andriot:2018wzk,Dvali:2018fqu,Achucarro:2018vey,Garg:2018reu,Kehagias:2018uem,Chiang:2018jdg,Heisenberg:2018yae,Kinney:2018nny,
Cicoli:2018kdo,Kachru:2018aqn,Akrami:2018ylq,Marsh:2018kub,Danielsson:2018qpa}), which suggest that it is not possible to construct metastable de Sitter vacua in a controlled way within string theory. As a corollary, if string theory is the correct high-energy description, the current period of accelerated expansion should be sourced by a quintessence field, e.g. through slowly rolling moduli field which naturally arise in string compactification scenarios. If future long baseline experiments should find the neutrino mass ordering to be inverted, this scenario would naturally be put under pressure, with extremely interesting implications concerning viable high-energy theories.

Finally, we quantify the preference for the normal ordering within NPDDE models in terms of probability odds ($p_{\rm NO}:p_{\rm IO}$). We adopt the methodology outlined in Sec.~\ref{sec:method}. For the NPDDE model, where $w(z)\geq-1$, we find that the normal ordering is mildly preferred with posterior odds $\sim$2:1 for the \textit{base} dataset and $\sim$3:1 for the \textit{pol} dataset.

We compare these figures to those obtained in  the generic $w_0w_a$CDM model. In this case, we find no preference for any of the two orderings for both the \textit{base} and \textit{pol} datasets (posterior odds of $\sim$1:1). When assuming the standard $\lcdm$ cosmological scenario, we find a mild preference for normal ordering of $\sim$2:1 for both the \textit{base} and \textit{pol} dataset combinations.

Finally, in Appendix~\ref{app:profile}, we provide an alternative approach to quantify the preference for the normal ordering. This alternative approach is based on the Akaike Information Criterion (AIC) estimator for the relative quality of statistical models. The findings are qualitatively in agreement with those reported in this section.

\section{Summary and Discussion}
\label{sec:conclusion}

A dynamical dark energy (DDE) component driving cosmic acceleration provides an alternative to the cosmological constant. In this work, we have explored cosmological constraints on the sum of the three active neutrino masses $M_{\nu}$ within DDE models. We parametrize the dark energy equation of state (EoS) as a function of redshift $z$ through the usual CPL parametrization $w(z)=w_0+w_a z/(1+z)$. Furthermore, we impose the requirement that  the EoS satisfies $w(z)\geq-1$ throughout the expansion history. We refer to this class of models as non-phantom dynamical dark energy (NPDDE). We employ a combination of CMB, BAO and SNeIa measurements. We denote by \textit{base} the dataset combination not including CMB polarization data at small scales, and by \textit{pol} the dataset combination which includes these CMB polarization data.

The conclusions we reach are threefold:
\begin{itemize}
\item We find that the constraints on $M_\nu$ assuming a NPDDE model are slightly tighter than those obtained within the standard $\lcdm$ scenario. This is the opposite of what is found when a generic DDE model with EoS allowed to enter the region where $w(z)<-1$ ($w_0w_a$CDM model) is assumed. More in detail, we find 95\%~C.I. upper bounds of $\mnu<0.13\,\eV$ for the \textit{base} dataset and $\mnu<0.11\,\eV$ for the \textit{pol} dataset in a NPDDE model. These figures can be compared to the 95\%~C.I. upper bounds of $\mnu<0.16\,\eV$ for the \textit{base} dataset and $\mnu<0.13\,\eV$ for the \textit{pol} dataset in a $\lcdm$ model. For the $w_0w_a$CDM model, we find instead the 95\%~C.I. upper bounds of $\mnu<0.41\,\eV$ for the \textit{base} dataset and $\mnu<0.37\,\eV$ for the \textit{pol} dataset. We provide a thorough data-supported physical and statistical explanation of these results. The explanation is based on the effects of massive neutrinos and dark energy on the background cosmological evolution, as well as on the Bayesian statistical method adopted.

\item A DDE component with $w(z)\geq-1$ does not alleviate the tension between cosmological and direct measurements of $H_0$, contrary to what is found in dark energy models with arbitrary $w(z)$. We find that NPDDE models prefer lower values of $H_0$ than those inferred by direct measurements. We also show that the well known degeneracy between $H_0$ and $M_\nu$ is reduced within NPDDE models. We provide a thorough explanation of this finding.

\item We combine the results of the cosmological analysis with neutrino oscillation data, and quantify the statistical preference for one of the two neutrino mass orderings over the other. The constraints on $\mnu$ in NPDDE models correspond to probability odds of $\sim$2:1 in favour of normal ordering with respect to inverted ordering for the \textit{base} dataset combination, and $\sim$3:1 for the \textit{pol} dataset. These odds show a mild preference for normal ordering. If laboratory experiments determine that the neutrino mass ordering is inverted, and if the current cosmic acceleration is caused by a dynamical dark energy component, this component would likely be phantom [$w(z)<-1$], or at least have to cross the phantom divide at some point during the expansion history. The conclusion holds as long as we exclude non-standard scenarios either in the neutrino sector or in the gravity sector. Therefore, this result brings to light a perhaps unexpected connection between two at first glance seemingly disconnected fields: neutrino oscillation experiments and the nature of dark energy. In the near future, results from the former might be able to shed important light on the latter.
\end{itemize}

\begin{acknowledgments}
The authors would like to thank Per Andersen, Thejs Brinckmann, Eleonora Di Valentino, Enrique Fern\'andez Mart\'{\i}nez, Edvard M\"{o}rtsell, Eiichiro Komatsu, Massimiliano Lattanzi, Eric Linder, Matteo Martinelli, Savvas Nesseris, Lorenzo Sebastiani, Zachary Slepian, Alessandra Silvestri, Michael Turner, and Sergio Zerbini for enlightening discussions. This work is based on observations obtained with Planck (\href{http://www.esa.int/Planck}{www.esa.int/Planck}), an ESA science mission with instruments and contributions directly funded by ESA Member States, NASA, and Canada. We acknowledge use of the Planck Legacy Archive. We also acknowledge the use of computing facilities at NERSC. K.F. acknowledges support from DoE grant DE-SC0007859 at the University of Michigan as well as support from the Leinweber Center for Theoretical Physics. K.F., M.G., and S.V. acknowledge support by the Vetenskapsr\aa det (Swedish Research Council) through contract No. 638-2013-8993 and the Oskar Klein Centre for Cosmoparticle Physics. S.D. and A.G. acknowledge support by the Vetenskapsr\aa det, the Swedish Space Board, and the K \& A Wallenberg foundation. O.M. is supported by PROMETEO II/2014/050, by the Spanish Grant FPA2014--57816-P of the MINECO, by the MINECO Grant SEV-2014-0398 and by the European Union’s Horizon 2020 research and innovation programme under the Marie Sk\l odowska-Curie grant agreements 690575 and 674896. O.M. would like to thank the Fermilab Theoretical Physics Department for its hospitality.

{\it \textbf{Note---}} After our paper appeared on the arXiv, the work~\cite{Durrive:2018quo} appeared which considers also neutrino mass bounds within quintessence models. Their work differs from ours in the parametrizations adopted for $w(z)$. Moreover, after our paper appeared on the arXiv, two further works~\cite{Choudhury:2018byy,Choudhury:2018adz} confirmed our result that the bounds on $M_{\nu}$ in non-phantom dynamical dark energy models are tighter than those obtained in $\Lambda$CDM. All codes, chains, and scripts used to produce the results and plots of this work will be made publicly available at \href{https://github.com/sunnyvagnozzi/NPDDE}{\texttt{github.com/sunnyvagnozzi/NPDDE}} after acceptance of the paper in a journal.
\end{acknowledgments}

\vskip 0.3 cm

\appendix

\section{Including the neutrino oscillations likelihood}
\label{app:oscillations}

Throughout this work, we have imposed a top-hat prior on $M_{\nu}$ of $M_{\nu}\geq0\,{\rm eV}$. That is, we allowed values of $M_{\nu}$ below the minimum value set by oscillation experiments of $0.06\,{\rm eV}$. The rationale behind this choice, as we outlined in Sec.~\ref{sec:method}, was threefold:
\begin{itemize}
\item To obtain a bound relying \textit{exclusively} on cosmological data.
\item To remain open to the possibility of models with non-standard neutrino interactions leading to a neutrino energy density which is either vanishing or lower than the expectation in $\Lambda$CDM (e.g.~\cite{Beacom:2004yd}): at the level of cosmological data these effects can be phenomenologically captured by considering values of $M_{\nu}$ below the lower bound set by oscillation experiments.
\item To provide an (in)consistency test for DE models where the upper bound on $M_{\nu}$ ends up lying significantly below the lower bound set by oscillation experiments. While this possibility has not been realized in our work due to insufficient sensitivity, it might be realized in the near-future thanks to dramatic improvements in the sensitivity of future CMB, large-scale structure, and supernovae distance measurements data sets.
\end{itemize}
Moreover, as also explained in Sec.~\ref{sec:method}, the positivity of $M_{\nu}$ is the only genuine a priori information present in the problem, whereas the information that $M_{\nu} \geq 0.06\,{\rm eV}$ is not truly a priori, but rather a posteriori of oscillation experiments. Therefore, the formally correct way of incorporating such information is, in fact, by including the neutrino oscillations likelihood. In addition, as discussed in Sec.~\ref{sec:method}, as far as upper limits on $M_{\nu}$ (which are the subject of this work) are concerned, the oscillations likelihood can be to zeroth approximation included as a sharp cutoff at $M_{\nu}=0.06\,{\rm eV}$, because the uncertainty to which $M_{\nu,\mathrm{min}}\simeq0.06\, {\rm eV}$ is subject is extremely tiny (but an uncertainty is nonetheless present, whereas the physical lower bound $M_{\nu} \geq 0\,{\rm eV}$ is instead subject o no uncertainty). Let us first discuss this simplified case where the oscillations likelihood is simply included as a sharp cutoff in the $M_{\nu}$ prior, before discussing a more physical, but still simple, way of including the oscillations likelihood.

One might at this point wonder whether or not our results are dependent on the choice of prior: $M_{\nu} \geq 0\,{\rm eV}$ versus $M_{\nu} \geq 0.06\,{\rm eV}$. In fact, the specific bounds on $M_{\nu}$ within a given model (in this case, $w_0w_a$CDM, $\Lambda$CDM, and NPDDE) \textit{are} certainly affected by the choice of prior. Nonetheless, it is easy to show that if the prior on $M_{\nu}$ is chosen to be flat even when the lower bound from oscillation experiments is enforced, then as a consequence of Bayes' theorem the key result of our paper remains unchanged. That is, the constraints on $M_{\nu}$ in NPDDE models remain tighter than those obtained in $\Lambda$CDM, even when the lower limit of $M_{\nu}\geq0.06\,{\rm eV}$ is enforced.

Let us denote by $\boldsymbol{x}$ our data and by $\boldsymbol{\theta}$ the set of cosmological parameters excluding $M_{\nu}$. Let us further denote by ${\cal L}(\boldsymbol{x} \vert M_{\nu},\,\boldsymbol{\theta})$ our likelihood, and by $\pi(M_{\nu})$ and $\pi(\boldsymbol{\theta})$ the prior distributions on $M_{\nu}$ and $\boldsymbol{\theta}$ respectively. Note that we are implicitly assuming that the prior on $M_{\nu}$ can be factorized from the prior on the other cosmological parameters, an assumption which is realized. From Bayes' theorem we know that the posterior distribution of $M_{\nu}$ given the data, $p(M_{\nu} \vert \boldsymbol{x})$, is given by the following:
\begin{eqnarray}
p(M_{\nu} \vert \boldsymbol{x}) \propto \int d\boldsymbol{\theta}\ {\cal L}(\boldsymbol{x} \vert M_{\nu},\,\boldsymbol{\theta})\pi(M_{\nu})\pi(\boldsymbol{\theta}) \, .
\label{bayes}
\end{eqnarray}
Assuming we keep a flat prior on $M_{\nu}$, the only effect of imposing the lower limit from oscillation experiments is to cut $\pi(M_{\nu})$ at $0.06\,{\rm eV}$ instead of $0\,{\rm eV}$. From Eq.~(\ref{bayes}), we see that the result of this operation would be to shift the posterior of $M_{\nu}$ to higher values: this will affect all quantities computed from the distribution, such as the mean and the 95\%~C.I. upper bound, both of which would increase, hence leading to broader constraints.

However, in our work we are not interested in the bounds on $M_{\nu}$ \textit{per se}. The purpose of our work is to examine how the upper limits on $M_{\nu}$ change when moving from $\Lambda$CDM to NPDDE. In Fig.~\ref{mnu}, we showed how the posterior of $M_{\nu}$ obtained assuming the NPDDE model is shifted to lower values compared to the one obtained assuming the $\Lambda$CDM model. From Eq.~(\ref{bayes}), it is easy to see how this fact continues to be true even when the lower limit of $0.06\,{\rm eV}$ set by oscillation experiments is imposed. Therefore, we expect as a consequence of Bayes' theorem that the upper limits on $M_{\nu}$ in NPDDE models will still be tighter than those obtained in $\Lambda$CDM regardless of whether a prior of $M_{\nu}\geq0\,{\rm eV}$ or $M_{\nu}\geq0.06\,{\rm eV}$ is chosen.

To confirm the above statement explicitly, we recompute the posteriors and upper limits on $M_{\nu}$ obtained in Sec.~\ref{sec:results}, this time imposing the lower limit set by oscillation experiments. We recomputed the bounds only for the $\Lambda$CDM and NPDDE models (leaving aside $w_0w_a$CDM, since it is not important for our conclusions), and only for the \textit{base} dataset (since the \textit{pol} dataset leads to identical conclusions). For the $\Lambda$CDM case, we find that the 95\%~C.I. upper bound of $0.16\,{\rm eV}$ degrades to $0.19\,{\rm eV}$ when imposing $M_{\nu}\geq0.06\,{\rm eV}$. Similarly, when considering the NPDDE model, the limit degrades from $0.13\,{\rm eV}$ to $0.17\,{\rm eV}$. However, we see that in both cases the upper limit obtained assuming NPDDE is tighter than that obtained assuming $\Lambda$CDM, confirming the conclusion we reached previously on the basis of Bayes' theorem.

Recall also that we computed the posterior odds for normal ordering versus inverted ordering using the methodology of~\cite{Hannestad:2016fog} outlined in Sec.~\ref{sec:method}. We wish to clarify that the posterior odds computed in this way are not affected by the choice of prior on $M_{\nu}$. The reason is that, in Eq.~(\ref{eq:evidence}), the likelihood of the cosmological data ${\cal L}(D \vert m_0,O)$ has been rewritten in terms of the mass of the lightest neutrino mass eigenstate $m_0$ rather than the total neutrino mass $M_{\nu}$. The relation between $M_{\nu}$ involves the squared mass-splittings measured by oscillation experiments. Therefore, the methodology adopted factors in, by construction, the information concerning the lower limits on $M_{\nu}$ set for the normal and inverted orderings. As a result, this methodology automatically ignores the region of the $M_{\nu}$ posterior which lies below $0.06\,{\rm eV}$. In fact, it is easy to show that in the low-mass region of $M_{\nu}$ parameter space favoured by data, $M_{\nu}\lesssim 0.15\,{\rm eV}$, the posterior odds for normal versus inverted ordering calculated using Eq.~(\ref{eq:evidence}) are well approximated by the following~\cite{Vagnozzi:2017ovm}:
\begin{eqnarray}
\frac{p_N}{p_O} \approx \frac{\int_{0.06\,{\rm eV}}^{\infty}dM_{\nu}\,p(M_{\nu} \vert \boldsymbol{x})}{\int_{0.10\,{\rm eV}}^{\infty}dM_{\nu}\,p(M_{\nu} \vert \boldsymbol{x})} \, ,
\label{evidence2}
\end{eqnarray}
where $p(M_{\nu} \vert \boldsymbol{x})$ is the posterior of $M_{\nu}$. The form of Eq.~(\ref{evidence2}) shows how the information on the lower limits of $M_{\nu}$ for both the normal and inverted ordering enters by construction in the methodology adopted.

So far, we considered the extremely simplified case where the information from oscillation experiments is included as a sharp cutoff in the $M_{\nu}$ prior. In fact, the minimum value allowed by oscillation experiments does not come without uncertainty. Using the best-fit values and $1\sigma$ intervals from~\cite{deSalas:2017kay} for the two mass-squared splittings assuming the normal ordering, we find $M_{\nu\,,\min} \approx 0.0589 \pm 0.0005\,{\rm eV}$. Therefore, we can to first approximation treat the oscillations likelihood as being a constant down to $0.0589\,{\rm eV}$, and as a truncated Gaussian centered at $0.0589\,{\rm eV}$ and with width $0.0005\,{\rm eV}$ below. If we simply included the oscillations likelihood as a sharp cutoff, the result on the $M_{\nu}$ posterior would be a sharp drop at $0.06\,{\rm eV}$, which is in a sense ``unphysical'' because this cutoff is actually induced by the value of $M_{\nu\,,\min}$ inferred from oscillation experiments with some uncertaintly, and not by a physical lower bound which comes without uncertainty (such as $M_{\nu} \geq 0\,{\rm eV}$). Nonetheless, since $M_{\nu\,,\min}$ is known to better than $\approx 0.8\%$ precision, it is perfectly reasonable to expect that the impact on the upper limits on $M_{\nu}$ of using our smeared Gaussian approximation versus a sharp cutoff is going to be negligible at best (which we later confirm).

\begin{figure}[!h]
\includegraphics[width=0.9\linewidth]{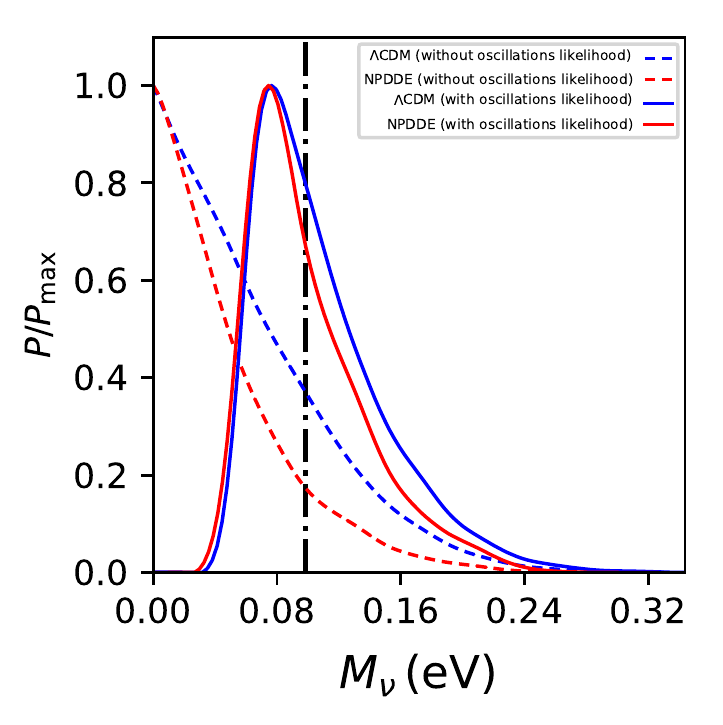}
\caption{One-dimensional posterior probabilities of the sum of the three active neutrino masses $M_{\nu}$ (in eV) for two models: the $\lcdm$ case (in blue), and the non-phantom dynamical dark energy (NPDDE) model with $w(z)\geq-1$ (in red), considering both the case when the oscillations likelihood is not included (dashed) and the case where it is included (solid). We approximate the oscillations likelihood as being a constant for $M_{\nu} \geq 0.0589\,{\rm eV}$ and a truncated Gaussian with width $0.0005\,{\rm eV}$ for $M_{\nu}<0.0589\,{\rm eV}$. The vertical black dotted-dashed line corresponds to the minimal mass of $M_{\nu,\min} \approx 0.1\, {\rm eV}$ allowed by neutrino oscillation data within the inverted ordering. We visually see that the dashed/solid red curves are always shifted to lower values of $M_{\nu}$ compared to the respective dashed/solid blue curves. Therefore, the upper limits on $M_{\nu}$ are always tighter in the NPDDE model compared to the $\Lambda$CDM one, independently of whether or not the oscillations likelihood is included.}
\label{fig6}
\end{figure}

In Fig.~\ref{fig6}, we show the posteriors we obtained for the $\Lambda$CDM (blue) and NPDDE (red) models, for both the case where the oscillations likelihood is not included (dashed) and the case where it is included (solid). We immediately notice two things. The first is that, as expected, posterior drops very sharply below $0.0589\,{\rm eV}$, signalling that the sharp cutoff approximation to the oscillations likelihood is in fact a reasonable approximation, given the extremely tiny uncertainty on $M_{\nu\,,\min}$. In fact, we find that we recover the upper limits we computed previously from the sharp cutoff approximation ($0.17\,{\rm eV}$ for the NPDDE model and $0.19\,{\rm eV}$ for the $\Lambda$CDM model). The second observation is that, independently of whether or not the oscillations likelihood is included, the NPDDE posterior is always shifted to lower values of $M_{\nu}$ compared to the $\Lambda$CDM one, showing that the main conclusions of our paper are stable against the inclusion of the oscillations likelihood.

To conclude, we summarize the findings of this Appendix. The key conclusion of our work, namely the fact that the upper limits on $M_{\nu}$ are tighter in NPDDE models compared to $\Lambda$CDM, persists even when the oscillations likelihood is included. Approximating the oscillations likelihood as a sharp cutoff in $M_{\nu}$ (which we have argued is a reasonable zeroth approximation given the very tiny uncertainty on $M_{\nu\,,\min}$), we have shown how this result follows simply from Bayes' theorem. Using a more realistic approximation to the oscillations likelihood (treated as a truncated Gaussian), we show the impact of including this likelihood on the $M_{\nu}$ posterior in Fig.~\ref{fig6}. Since the methodology adopted to compute the posterior odds of normal versus inverted ordering (see Sec.~\ref{sec:method} and~\cite{Hannestad:2016fog}) by construction takes into account the lower limits on $M_{\nu}$ coming from oscillation experiments for both normal and inverted ordering, the results obtained in Sec.~\ref{subsec:determination} are unaffected by whether or not the lower limit of $M_{\nu}\geq0.06\,{\rm eV}$ is enforced. Therefore, in NPDDE models (and using the \textit{base} dataset), the preference for normal versus inverted ordering is $\sim 2:1$.

\section{Estimating the preference for the normal ordering through the Akaike information criterion}
\label{app:profile}
We complete the analysis of this work by quantifying the preference for one neutrino mass ordering over the other using an alternative statistical method based on the Akaike information criterion (AIC)~\cite{akaike}. The AIC is a statistical indicator which estimates the relative statistical quality of different models. We use the AIC to estimate the preference for the normal ordering (NO) against the inverted ordering (IO). For a model with $k$ parameters and log-likelihood $\ln({\cal L}) = -\chi^2/2$, the AIC is given by:
\begin{eqnarray}
{\rm AIC} = 2k+\min(\chi^2) \, ,
\end{eqnarray}
where $\min(\chi^2)$ denotes the minimum value of the $\chi^2$ for the model. The difference between the AICs of two models, $\Delta {\rm AIC}$, estimates the relative quality of one model against the other. In particular, the model with the lowest AIC is to be considered statistically preferred.

As in Sec.~\ref{sec:results}, we combine results from cosmology and neutrino oscillation experiments to quantify the preference for the normal ordering. We obtain the posterior probability distribution of $M_{\nu}$ from the cosmological analysis, and interpret this posterior as a likelihood for the cosmological dataset (as done in~\cite{Hannestad:2016fog, Vagnozzi:2017ovm}). From oscillation measurements, we take the one-dimensional $\chi^2$ projections for the solar and atmospheric mass splittings computed separately for NO and IO, as provided by NuFIT 3.0 (2016)~\cite{Esteban:2016qun}. We then compute the global $\min(\chi^2)$ for both NO and IO in light of the combination of cosmological and oscillation data. 

The number of parameters $k$ is the same in the two scenarios, therefore $\Delta\mathrm{AIC}=\Delta\min(\chi^2)$. For NPDDE models, we find $\Delta\mathrm{AIC}=\mathrm{AIC_{IO}}-\mathrm{AIC_{NO}}=3.4$ for the \textit{base} dataset and $\Delta\mathrm{AIC}=3.7$ for the \textit{pol} dataset. These values show a mild preference for the NO model, using the scale provided by~\cite{kass}. In $\lcdm$, we find $\Delta\mathrm{AIC}=2.7$ for the \textit{base} dataset and $\Delta\mathrm{AIC}=3.1$ for the \textit{pol} dataset. In this case, the preference for NO is even milder. Finally, in generic DDE models with arbitrary EoS, we find $\Delta\mathrm{AIC}=1.9$ for the \textit{base} dataset and $\Delta\mathrm{AIC}=2.0$ for the \textit{pol} dataset. We interpret these values as a minimal preference for NO. The results obtained using the AIC method are qualitatively in agreement with those obtained performing a Bayesian model comparison and reported in Sec.~\ref{sec:results}. Both indicate that current cosmological data, when interpreted in light of a NPDDE model, show a mild preference for the NO.

\bibliographystyle{JHEP}
\bibliography{de.bib}

\end{document}